\documentclass[twocolumn,prd,superscriptaddress,showpacs,amsmath,amssymb]{revtex4-1}

\usepackage{graphicx}
\usepackage{dcolumn}
\usepackage{bm}
\usepackage{multirow}

\begin{document}

\title{Search for An Annual Modulation in \\ Three Years of CoGeNT
Dark Matter Detector Data}

\def\PNNL{Pacific Northwest Laboratory, Richland, WA 99352, USA}
\def\Duke{Department of Physics, Duke University, Durham, NC 27708,
USA}
\def\Tunl{Triangle Universities Nuclear Laboratory, Durham, North
Carolina, 27708, USA}
\def\UC{Kavli Institute for Cosmological Physics and Enrico Fermi
Institute, University of Chicago, Chicago, IL 60637, USA}
\def\Canberra{CANBERRA Industries, Meriden, CT 06450, USA}
\def\UW{Center for Experimental Nuclear Physics and Astrophysics and
Department of Physics, University of Washington, Seattle, WA 98195,
USA}

\author{C.E.~Aalseth} \affiliation{\PNNL}
\author{P.S.~Barbeau} \affiliation{\Duke}\affiliation{\Tunl} 
\author{J.~Colaresi} \affiliation{\Canberra}
\author{J.I.~Collar} \email{Electronic address:
collar@uchicago.edu}\affiliation{\UC}
\author{J.~Diaz Leon} \affiliation{\UW}
\author{J.E.~Fast} \affiliation{\PNNL}
\author{N.E.~Fields} \affiliation{\UC}
\author{T.W.~Hossbach} \affiliation{\PNNL} 
\author{A.~Knecht} \affiliation{\UW}
\author{M.S.~Kos} \affiliation{\PNNL}
\author{M.G.~Marino} \altaffiliation{Present address: Physics
Department, Technische Universit\"at M\"unchen, Munich,
Germany}\affiliation{\UW}
\author{H.S.~Miley} \affiliation{\PNNL}
\author{M.L.~Miller} \altaffiliation{Present address: Cloudant - West
Coast, 209 1/2 1$^{st}$ Ave S,  Seattle, WA 98104}\affiliation{\UW}
\author{J.L.~Orrell}  \email{Electronic address:
john.orrell@pnnl.gov}\affiliation{\PNNL}
\author{K.M.~Yocum} \affiliation{\Canberra}

\collaboration{CoGeNT Collaboration}

\date{\today}

\begin{abstract}
Weakly Interacting Massive Particles (WIMPs) are well-established dark matter candidates. WIMP interactions with sensitive detectors are expected to display a characteristic annual modulation in rate. We release a dataset spanning 3.4 years of operation from a low-background germanium detector, designed to search for this signature. A previously reported modulation persists, concentrated in a region of the energy spectrum populated by an exponential excess of unknown origin. Its phase and period agree with phenomenological expectations, but its amplitude is a factor $\sim$4-7 larger than predicted for a standard WIMP galactic halo. We consider the possibility of a non-Maxwellian local halo velocity distribution as a plausible explanation, able to help reconcile recently reported WIMP search anomalies.
\end{abstract}



\maketitle

\section*{Introduction}

The denomination "Weakly Interacting Massive Particle" (WIMP) encompasses a large family of hypothetical heavy neutral particles having weak-scale interactions. Presently favored extensions of the standard model of particle interactions naturally generate WIMP candidates with properties able to account for the bulk of the missing mass of the universe, the so-called "dark matter", a term coined by F. Zwicky in 1933 \cite{zwicky}. The most recent precision-cosmology measurements \cite{planck,wmap} set the universal abundance of this type of invisible matter, unable to emit or absorb light, at a factor of five to six times that for visible (baryonic) matter. WIMPs thermally produced during early epochs, with masses in the broad range GeV-TeV, have annihilation cross-sections that naturally lead to the present abundance of dark matter, an observation often referred to as "the WIMP miracle" \cite{miracle}. Over the past few decades, abundant astronomical evidence has accumulated pointing at the existence of vast dark galactic haloes surrounding the luminous portion of most galaxies. This evidence has been most recently backed-up by numerical simulations of galaxy formation and evolution \cite{frenk}. A large number of underground experiments are currently searching for indications of characteristic WIMP interactions in a variety of target detector materials \cite{rick}. 

The CoGeNT detector presently operating at the Soudan Underground Laboratory (SUL) \cite{longcogent} was designed to investigate the hypothesis, put forward in 2004 \cite{bottino,gondolo1}, that low-mass ($m_{\chi}\!<\!10$~GeV/c$^{2}$) WIMPs \cite{ref1,ref2,ref3,ref4} might be the source of the annual modulation effect observed by the DAMA/LIBRA arrays of low-background NaI[Tl] scintillators \cite{DAMA}. This effect had been previously predicted to arise from the motion of the Earth-Sun system relative to a WIMP dark matter galactic halo engulfing our galaxy \cite{andrzej}. This expected annual modulation in interaction rate is arguably the most prominent signature presently available for the positive identification of WIMPs. 

CoGeNT employs p-type point contact (PPC) germanium detectors \cite{jcap} for this purpose. PPCs feature very low electronic noise, allowing them to detect the small (sub-keVee) energy depositions that would be produced by elastic scattering of low-mass WIMPs off nuclei \cite{witten,wasserman}. An early CoGeNT prototype, operated in a shallow underground site \cite{Aal08,Aal08err}, excluded the last region of low-mass WIMP parameter space (WIMP mass, nuclear scattering cross-section) able to account for the DAMA/LIBRA modulation, under the premise of a local galactic halo described by an isotropic Maxwellian velocity distribution, the simplest approximation to the motion of these particles \cite{smith}. 

A search for an annual modulation induced by dark matter particle interactions, a subtle few-percent variation in the low-energy interaction rate, is a challenging endeavor. It requires excellent long-term stability in the entire detector system. We have recently discussed these difficulties within the context of our experiment, concluding that the necessary conditions seem present \cite{longcogent}. In this same publication, we described an improved approach at discriminating against backgrounds affecting the surface layers of the detector. This rejection is based on a measurement of the rise-time ($t_{10-90}$) of preamplifier pulses, an effect previously introduced in \cite{Aal11,Aal11b}. Surface events are observed to produce considerably larger values of $t_{10-90}$ than energy depositions taking place within the bulk of the germanium crystal (Fig.\ 1). The long exposure accumulated by the PPC at SUL has allowed us to characterize the rise-time distributions of these two populations. These distributions can be described by log-normal functions, a choice justified by simulations of the effect of electronic noise on rise-time determination, and electronic pulser calibrations \cite{longcogent}. We observe a monotonic evolution of the fit parameters (amplitude, mean, and variance of both distributions) with energy, following trends predicted by the simulations (Fig.\ 1, \cite{longcogent}).

Operation of the PPC at SUL was interrupted by a fire in the vertical shaft access to the laboratory. We accumulated fifteen months of data previous to this incident, reporting on the observation of a possible annual modulation in that dataset \cite{Aal11b}. Operation restarted on 7 June 2011 following a three month downtime. The detector has taken data continuously since this breach, with no significant change in performance (Fig.\ 2). The presently examined cumulative dataset ends on 23 April 2013, spanning 1,237 days, of which 1,129 were live. The purpose of this communication is to illustrate one possible form of analysis for the accumulated dataset, accompanying its public release \cite{databases} in time-stamp, energy, and rise-time format, for each individual event passing quality cuts.

 \begin{figure}[!htbp]
\includegraphics[width=0.3\textheight]{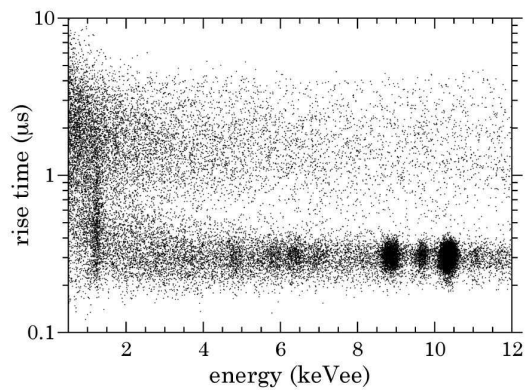}
\caption{\label{} Rise-time of preamplifier traces plotted against
energy deposition, for each event passing quality cuts \protect\cite{longcogent}  in the present dataset. Two distinct
families of events are visible, surface events (top band), and events
from the bulk of the crystal (lower band). Cosmogenic peaks populate
the lower band. A large accumulated exposure has allowed the
characterization of the demarcation of these two families as a
function of energy \cite{longcogent}.}
\end{figure}

\begin{figure}[!htbp]
\includegraphics[width=0.3\textheight]{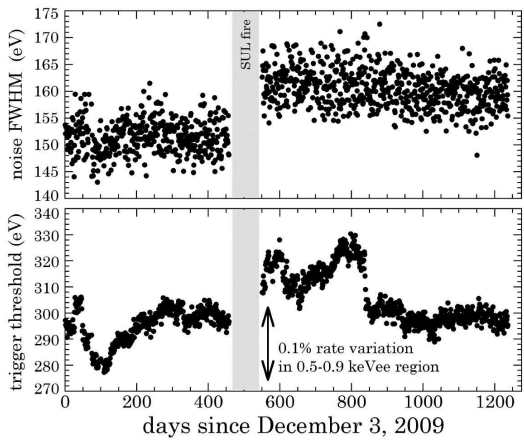}
\caption{\label{} Daily average electronic noise and trigger
threshold stability \protect\cite{Aal11b} in the CoGeNT PPC detector
at SUL. The change in electronic noise following the thermal cycle
imposed by the SUL fire is too small to measurably impact energy
threshold or resolution. Negligible fluctuations in trigger threshold
level are expected from the $\sim1^{\circ}$C temperature stability of
the data acquisition electronics \protect\cite{longcogent}.}
\end{figure}

\section*{Data Analysis}

Fig.\ 3 displays the $t_{10-90}$ distributions for events passing software cuts against microphonics and other spurious pulses \cite{longcogent}, for two energy regions. The first, spanning the range 0.5-2.0 keVee, goes from analysis threshold to the approximate endpoint of an exponential excess of bulk events with presently unknown origin \cite{longcogent,Aal11,Aal11b}. If this excess is identified with a possible WIMP signal, the corresponding best-fit WIMP mass would be $m_{\chi}\sim$8 GeV/c$^{2}$ \cite{longcogent,Aal11,Aal11b}. The second energy region goes from 2.0 keVee up to 4.5 keVee. Above 4.5 keVee the energy spectrum is encumbered by decaying peaks corresponding to K-shell electron capture (EC) following cosmogenic activation of germanium \cite{Aal11, Aal11b}. Log-normal fits to fast (bulk) and slow (surface) event populations are shown in Fig.\ 3, as in \cite{longcogent}.

\begin{figure}[!htbp]
\includegraphics[width=0.3\textheight]{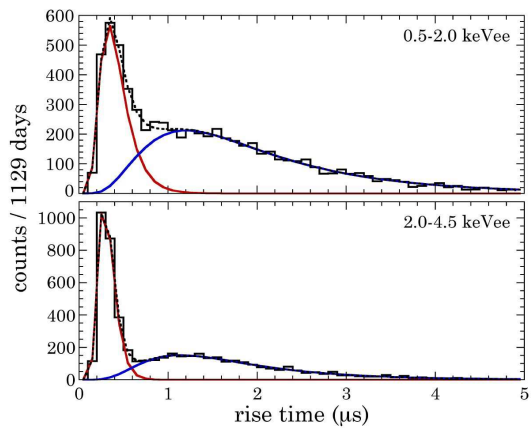}
\caption{\label{} Rise time distributions (histograms) of the two energy regions
examined in the present analysis. Log-normal fits to fast (bulk)
events are indicated in red, slow (surface) events in blue, and their
sum is a dotted line.}
\end{figure} 

 For the purposes of the present analysis, we further divide these two energy regions into ``pure" bulk and ``pure" surface events, separated at the approximate value of $t_{10-90}$ for which the probability of belonging to either the fast or slow signal populations is the same (0.7 $\mu$s for 0.5-2.0 keVee, and 0.6 $\mu$s for 2.0-4.5 keVee, Fig.\ 3). In doing so, the fast signal acceptance (SA) for the ``pure" bulk 0.5-2.0 keVee events is 92.2\%, whereas the slow signal background rejection (BR) is 93.1\%, leading to a $\sim$13.6\% contamination of this group by surface backgrounds. For the ``pure" surface 0.5-2.0 keVee group, the SA and BR values are exchanged, leading to an expected contamination by fast pulses of just 3.8\%. In the 2.0-4.5 keVee energy range these SA and BR are 98.6\% and 95.75\%, respectively, leading to a surface contamination of the ``pure" bulk group by 4.4\% of the events, and bulk contamination of the ``pure" surface group by 1.3\%. The expectation is that an annual modulation effect should appear only in the low-energy (0.5-2.0 keVee) ``pure" bulk group, if the exponential spectral excess there were to be associated with a low-mass WIMP signal. 
 
 Fig.\ 4 shows the temporal variation of event rates in these four groups of signals. The decay associated to cosmogenic activation in germanium and cryostat materials \cite{cosmo1,cosmo2} is evident in this long exposure. It is particularly apparent for the 0.5-2.0 keVee bulk group, which contains the majority of L-shell EC signals, readily observable as peaks in the energy spectrum \cite{Aal11, Aal11b}. Their contribution is calculable through monitoring of the higher-intensity K-shell EC peaks populating the 4.5-11.5 keVee energy range \cite{Aal11b}, and use of well-characterized L/K EC ratios \cite{Bahcall,bamby,bambyerr}. Corrections are applied to this calculated L-shell EC rate for the fraction of fast signals not passing the $t_{10-90}<0.7~ \mu$s cut, and combined trigger plus microphonic cut efficiency (dashed line in Fig.\ 3 of \cite{Aal11}). Robust fits to the decays of the K-shell peaks make the uncertainty in this calculation small, as indicated in the figure \cite{note}. The 0.5-2.0 keVee surface group is expected to contain a fraction of these L-shell EC signals, from contamination with fast pulses with $t_{10-90}>0.7 ~\mu$s, and through the calculable (but somewhat uncertain) fraction of L-shell EC episodes taking place in a $\sim\!1$ mm-thick transition surface layer \cite{Aal11b,longcogent}. In addition to these calculable contributions, all four groups of events are subject to the decaying continuum from cosmogenic radioisotopes in materials next to the crystal, and internal cosmogenics not proceeding through EC (e.g., the dominant $\beta^{+}$ branch in the decay of $^{68}$Ga, $^{3}$H, etc. \cite{cosmo1,cosmo2}). While all four groups exhibit comparably long-lived decays, these undefined additional processes do not necessarily affect them equally, i.e., their rate evolution can be described by allowing independent decay constants.  
 
 \begin{figure}[!htbp]
\includegraphics[width=0.32\textheight]{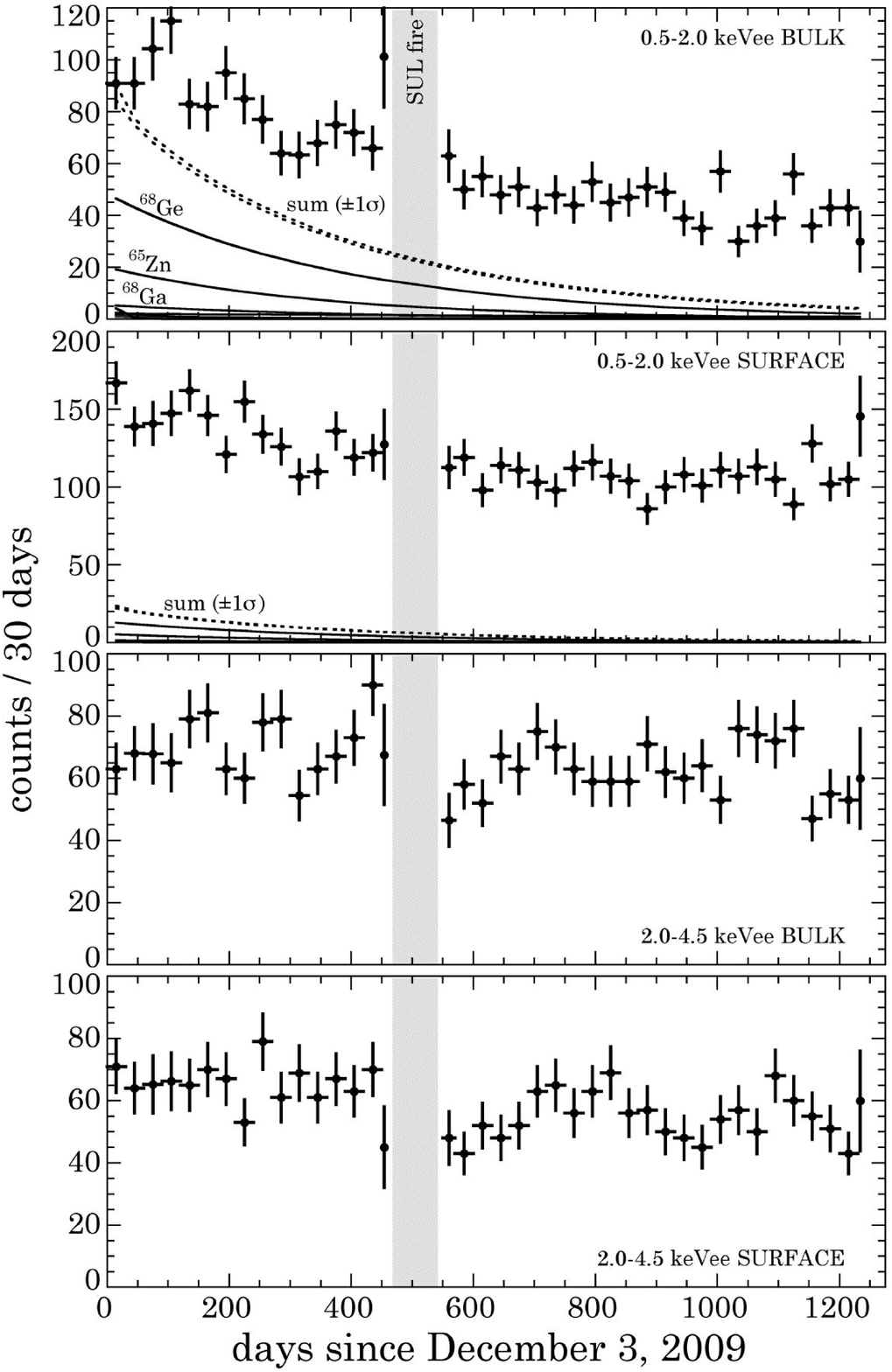}
\caption{\label{} Temporal evolution of the counting rate in the four
groups of signals considered. The calculated contributions from
L-shell EC are shown in the top two panels (see text). Dotted lines
indicate the $\pm1 \sigma$ uncertainty in their sum
\protect\cite{note}. Subdominant decays from $^{49}$V,  $^{51}$Cr,
$^{54}$Mn, $^{55}$Fe, $^{56-58}$Co and $^{71}$Ge are shown, but are
too small to be individually labelled. The contribution from
$^{56}$Ni is negligible, and that from $^{73}$As is removed by a
low-energy cut on time-coincident events \protect\cite{longcogent}. }
\end{figure}

Fig.\ 5 displays these rates after accounting for the decaying background components. The top panel represents the low-energy bulk group following subtraction of the calculated L-shell EC contribution. The rest of the panels show the residual rates in each group after subtraction of an exponentially-decaying background component with free half-life $T_{1/2}$, fitted to the rates in Fig.\ 4 as part of a model also containing a modulated component around a free constant rate, with fractional amplitude $S$, peak date $t_{max}$, and period $T$. The peak date (days since January 1st for the maximum) defines the phase of the modulation. Dotted lines show the best fits for unconstrained values of the modulation parameters, solid lines when $T=$ 365 days is imposed. We observe little change in the returned best fits for the 0.5-2.0 keVee bulk modulation parameters, regardless of if the L-shell EC contribution is calculated and removed, or if a free $T_{1/2}$ is allowed. In the first case a $T= 336\pm24$ days is found, in the second this is $T= 350\pm20$ days, both compatible with an annual modulation within uncertainties. Similarly, the peak date associated to $T=$ 365 days for this group of events is $t_{max}=102\pm47$ days. We note this is compatible with the $t_{max}=136\pm7$ days found for DAMA/LIBRA in the 2-4 keVee region of its spectrum where its modulation is maximal \cite{DAMA,kelso}. Best-fitted $T$ and $t_{max}$ for the other three groups of events appear at random values. Fits to these other three groups with $T=$ 365 days imposed do not favor the presence of a modulation (Fig.\ 5). We ascertain that significant power centered around  $T=$ 365 days appears only for the low-energy bulk group via a periodogram analysis (Fig.\ 6, \cite{period1,period2,note4}), taking binning precautions similar to those described in \cite{neal}.

\begin{figure}[!htbp]
\includegraphics[width=0.32\textheight]{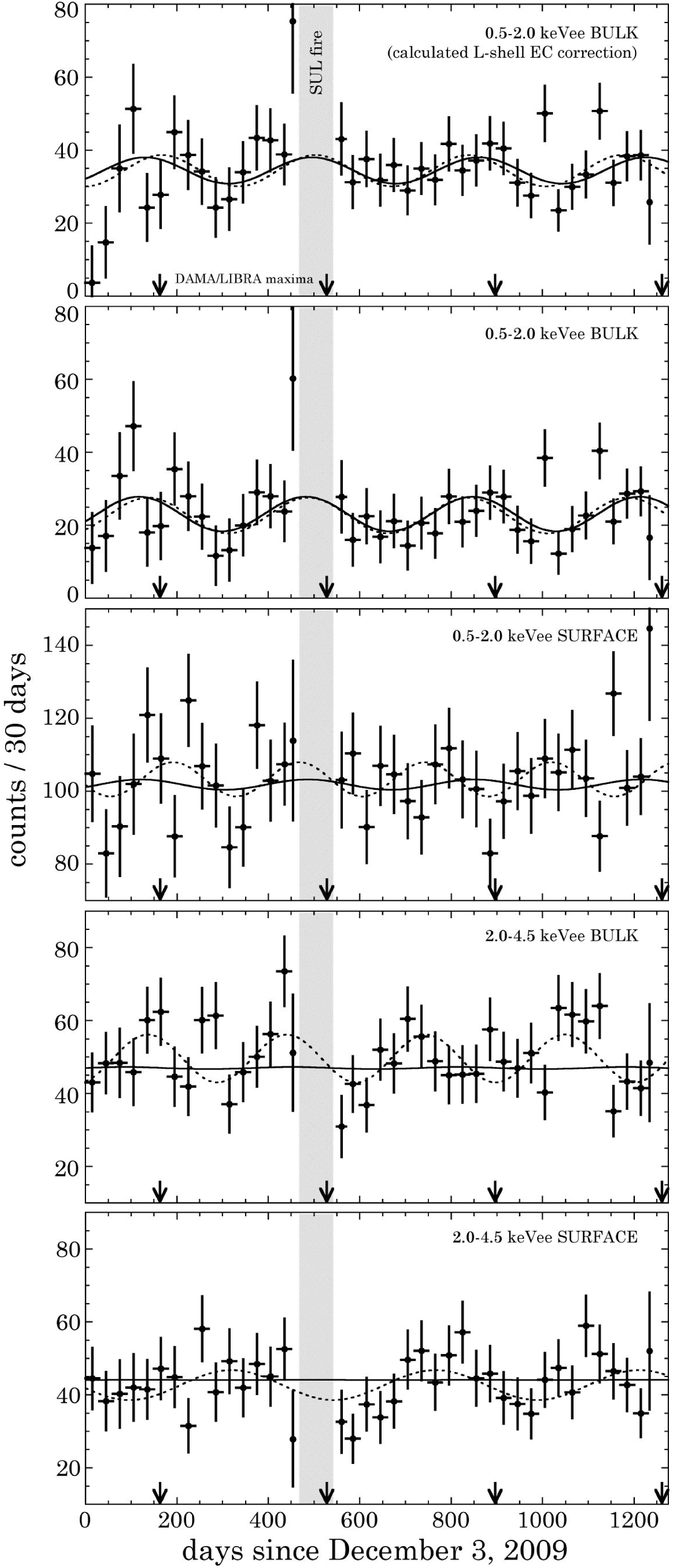}
\caption{\label{} Best-fit modulations for the four groups of events,
after accounting for decaying background components (see text).
Dotted lines and data points are for unconstrained modulations, solid
lines for an imposed annual period. Vertical arrows point at the
position of the DAMA/LIBRA modulation maxima \protect\cite{DAMA}. A
modulation compatible with a galactic dark halo is found exclusively
for bulk events, and only in the spectral region where a WIMP-like exponential 
excess of events is present.}
\end{figure}

\begin{figure}[!htbp]
\includegraphics[width=0.32\textheight]{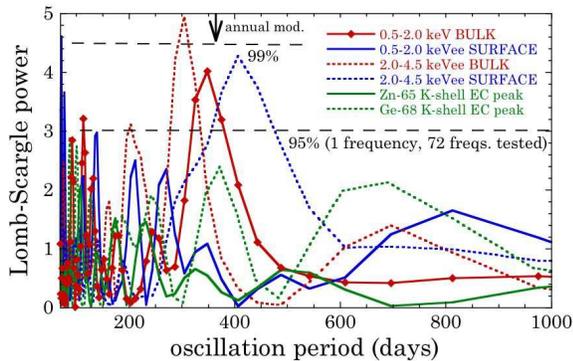}
\caption{\label{} Lomb-Scargle power \protect\cite{period1,period2}
for oscillations around the long-lived exponential background decays
in the four groups of events tested. Thirty-day time bins are used as in Figs.\ 4 and 5.
Power centered around $T\!=$ 365 days is found for the low-energy
bulk group only \protect\cite{note4}. We have not attempted a similar analysis method
claiming improved sensitivity to annual modulations
\protect\cite{freedman}. Green lines show this power for oscillations
of the two dominant K-shell EC signals around their exponential
decays. }
\end{figure} 

This straightforward treatment, which incorporates an improved discrimination against surface backgrounds compared to our previous analyses, confirms our earlier indication of an annual modulation in CoGeNT data \cite{Aal11b}, exclusively for the subset of events liable to contain a low-mass WIMP dark matter signal. Its significance is modest in the present unoptimized form of analysis: using the likelihood ratio method described in \cite{Aal11b} the hypothesis of an annual modulation being present in the low-energy bulk group is preferred to the null hypothesis (no modulation) at the $\sim2.2~ \sigma$ level \cite{note2,note3}. However, this frequentist approach does not take into consideration information from DAMA/LIBRA and other searches as a prior, specifically the potential relevance of the modulation amplitude favored by CoGeNT, a subject developed next. In this respect, we call attention to incipient applications of Bayesian methodology in this area \cite{arina1,arina2,arina3}. The remainder of this paper focuses on the possibility of using our observations to obtain a common phenomenological interpretation of recent intriguing results in direct searches for dark matter.

\section*{Discussion}

A best-fit value of $S\!=\!12.4(\pm5)$\% is observed for the low-energy bulk group when the L-shell EC contribution is  subtracted directly (top panel in Fig.\ 5). If a free $T_{1/2}$ is allowed (second panel in the figure), this becomes $S=21.7(\pm15)$\%. If the irreducible low-energy excess in the CoGeNT spectrum  is considered to be the response to a $m_{\chi}\sim$8 GeV/c$^{2}$ WIMP, it would account for $35$\% of the bulk events in the 0.5-2.0 keVee region, the rest arising from a flat component originating mainly in Compton scattering of gamma backgrounds (see discussion around Fig.\ 23 in \cite{longcogent}). This fraction is approximate, as it can change some with choice of background model, and of rise-time cuts leading to slight variations in the irreducible ``pure" bulk spectrum. This putative WIMP signal would then be oscillating with an annually-modulated fractional amplitude in the range between $\pm35$\% and $\pm62$\%. This is larger by a factor $\sim4\!-\!7$ than the $\pm9$\% expected for a WIMP of this mass in this germanium energy region, when the zeroth-order approximation of an isotropic Maxwellian halo is adopted \cite{smith}.

A growing consensus is that a Maxwellian description of the motion of dark matter particles in the local halo, the so-called standard halo model (SHM), is incomplete, as it excludes several expected halo components, e.g., tidal streams \cite{savage,natarajan,purcell}, debris flows \cite{kuhlen}, extragalactic components \cite{freese2,baushev}, a dark disk \cite{disk}, and other sources of anisotropies \cite{belli,green}. These can have a large effect on $S$ and $t_{max}$, in particular for low WIMP masses. For $m_{\chi}\sim10$ GeV/c$^{2}$, present detectors would be sensitive only to the highest velocities in the local halo velocity distribution \cite{freese3}. This can lead to enhanced fractional modulation amplitudes in a number of halo models \cite{purcell,kuhlen2,frandsen}. A useful quantity in this respect is the Earth's frame minimum WIMP velocity ($v_{min}$) that  can result in a nuclear recoil energy $E_{R}$, defined by $(v_{min}/c)^{2}\!=\!m_{N} E_{R}/2\mu^{2}$, where $m_{N}$ is the target's nuclear mass, and $\mu=m_{N}m_{\chi}/(m_{N}+m_{\chi})$ is the reduced mass of the WIMP/nucleus system \cite{purcell,kuhlen2,frandsen}.

Figs.\ 7 and 8 illustrate a point often missed: the DAMA/LIBRA region of interest (ROI) in WIMP parameter space (Fig.\ 8) is generated through the assumption of a fractional modulation amplitude corresponding to the SHM. Hence our original exclusion of this possibility (\cite{Aal08,Aal08err}, CoGeNT limit in Fig.\ 8), later reinforced by other experiments. Unlike the other searches generating anomalies and ROIs in Fig.\ 8, the DAMA/LIBRA spectrum shows no characteristic spectral excess next to threshold (Fig.\ 7), its single observable being the (absolute, not fractional) modulation amplitude. It is therefore possible to generate alternative DAMA/LIBRA ROIs for non-Maxwellian halo models \cite{belli,belli2}. 

\begin{figure}[!htbp]
\includegraphics[width=0.32\textheight]{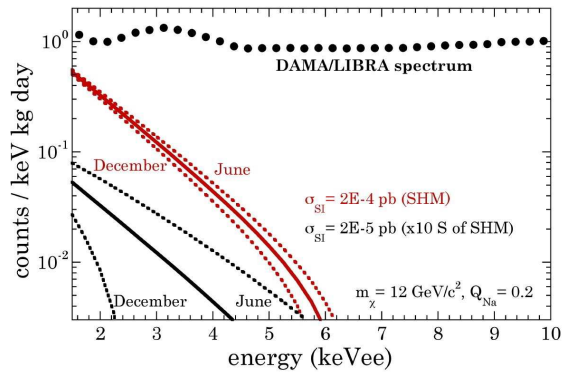}
\caption{\label{} DAMA/LIBRA energy spectrum. The peak around 3 keVee
has a known origin in $^{40}$K contamination. Two example average
WIMP signals, together their yearly extrema are shown. They differ in
scattering cross section $\sigma_{SI}$ and {\it fractional}  modulation amplitude
$S$, but generate the same observable, DAMA/LIBRA's {\it absolute}
modulation amplitude (units of counts per keV kg day). This
conceptual illustration does not include the spectral distortions
expected from specific deviations from the SHM
\protect\cite{freese3}, nor the effect of energy resolution.}
\end{figure}

Fig.\ 8 displays the effect on the DAMA/LIBRA ROI when a factor 6.8 larger fractional modulation than that predicted by the SHM is assumed, i.e., same as that found in the CoGeNT dataset (free $T_{1/2}$ case): a non-Maxwellian local halo favoring large values of $S$ is capable of reconciling the tension between DAMA/LIBRA and other recently reported anomalies, providing a coherent picture for these observations.  A separate possibility able to generate large fractional modulations is a WIMP with slightly smaller values of $m_{\chi}$ and larger values of $\sigma_{SI}$ relative to the ROIs in Fig.\ 8. This is a relevant region of parameter space \cite{andreas,model10}. This particle would be invisible (most of its recoils below detector thresholds) during winter, producing signals six months later, when the Earth-halo relative velocity is at a maximum. This possibility is ignored by analyses neglecting the large dispersion in NaI[Tl]  scintillation light yield caused by low-energy sodium recoils \cite{myq2}. 

\begin{figure}[!htbp]
\includegraphics[width=0.32\textheight]{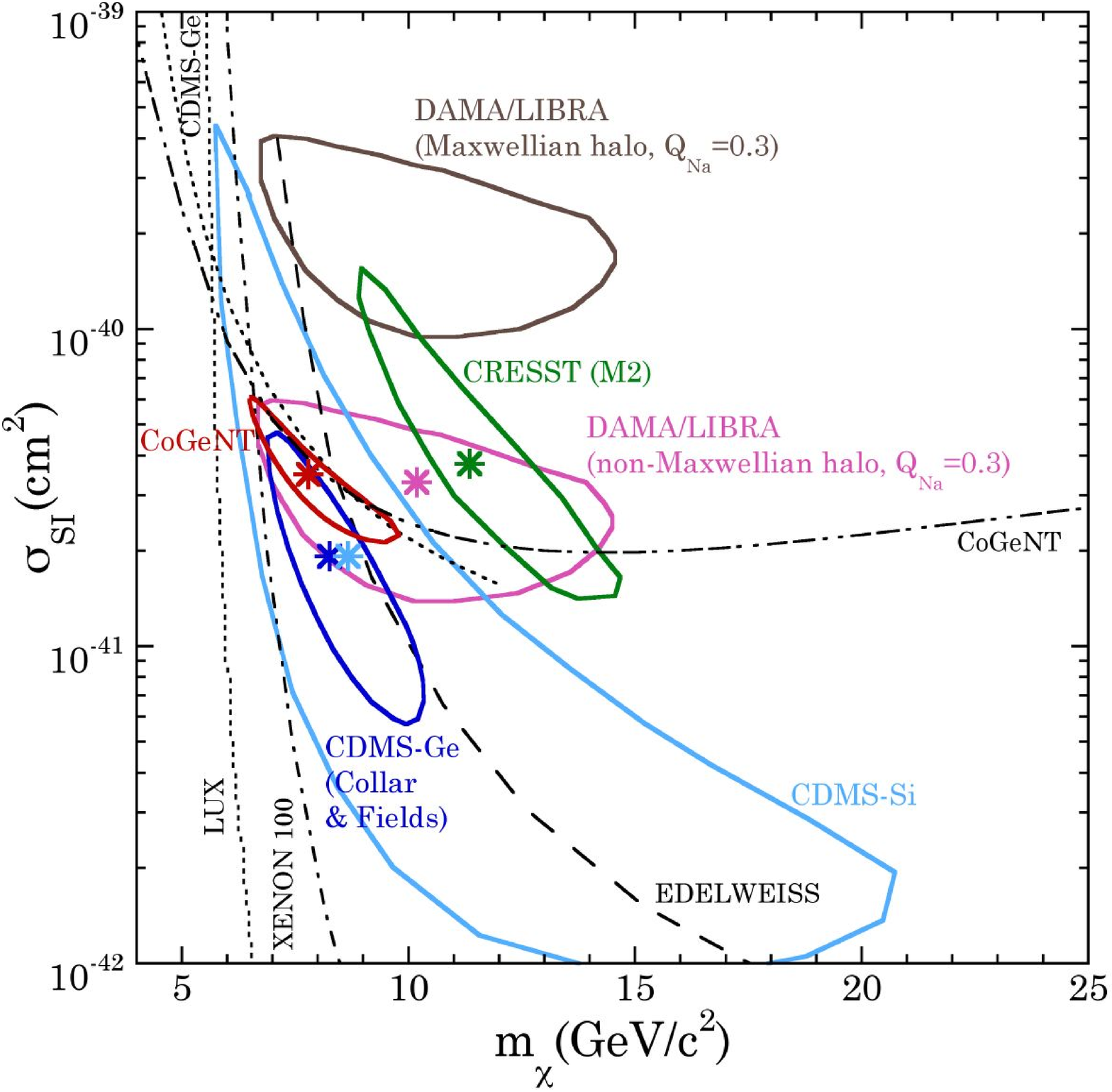}
\caption{\label{} Displacement towards lower spin-independent
scattering cross-section $\sigma_{SI}$ of the DAMA/LIBRA region of interest (ROI), if a
fractional modulation amplitude corresponding to that found for
CoGeNT data is assumed. The additional uncertainty generated by the value
of the sodium recoil quenching factor $Q_{Na}$ is discussed in the
text. ROIs for recent dark matter detector anomalies
\protect\cite{Aal11,Aal11b,longcogent,cresst,ourcdms,cdmssi} are
indicated by colored solid lines, their best-fits highlighted by
asterisks. These are 99\% C.L. regions, except for CRESST (95\%
C.L.). Limits from
\protect\cite{longcogent,cdmsge,edelweiss,xenon100,lux} are indicated by
dashed lines.  
A search for an annual modulation in CDMS germanium data \cite{cdmsmod} would be insensitive to up to a 100\% modulation amplitude in a possible CDMS-Ge signal \cite{ourcdms}. Liquid xenon (LUX, XENON-100) sensitivity  to $m_{\chi}\!<\!12$ GeV/c$^{2}$ is presently under test, using an $^{88}$Y/Be neutron source \cite{myq1}.
}
\end{figure}

Yet the situation is more complex than in the succinct description above. For instance, the most recent measurements of the quenching factor for low-energy sodium recoils in NaI[Tl] \cite{myq1,myq2}, point at a value $Q_{Na}\!\!\sim\!0.15$, considerably smaller than the traditionally adopted $Q_{Na}\!=\!0.3$. The CoGeNT quenching factor is comparatively well-established \cite{longcogent}. In the representation of Fig.\ 8, this would displace the centroid of the DAMA/LIBRA ROI from $m_{\chi}\!\sim\!10$ GeV/c$^{2}$ to $m_{\chi}\!\sim\!18$ GeV/c$^{2}$, i.e., away from other anomalies and into the excluded region of parameter space \cite{myq1}. It is nevertheless still possible to recover agreement by adopting recent methods aiming at removing astrophysical uncertainties from the interpretation of dark matter detector data (\cite{neal, kuhlen2,frandsen,paddy,paolograciela,paolograciela2,herrero,mao}, see specifically Fig.\ 6 in \cite{kelso}). A small $Q_{Na}$, if confirmed, would push DAMA/LIBRA's $v_{min}$ into the range $>$650 km/s, for which a large fractional modulation is expected from essentially all halo models \cite{purcell,kuhlen2,frandsen}, imposing a reduction in $\sigma_{SI}$ for the DAMA/LIBRA ROI similar to that shown in Fig.\ 8.  Also susceptible to  $Q_{Na}$ is the potential relevance of the small shift in  $t_{max}$ between CoGeNT and DAMA/LIBRA, given the dependence of $t_{max}$ on $E_{R}$ highlighted in \cite{kuhlen2} (a similar shift being already noticeable within DAMA/LIBRA data when examined for different $E_{R}$ regions \cite{DAMA,kelso,lisanti}). Gravitational focusing \cite{lisanti} can also affect the value of $t_{max}$ for different detectors. Finally, a lower  $\sigma_{SI}$ for DAMA/LIBRA, as suggested by the discussion above, would ease concerns about the ability of its spectrum to accommodate a dark matter signal \cite{kud,pradler}. 

\section*{Conclusions}

To summarize, following an improvement in our ability to separate bulk and surface signals, we report a continued preference for an annual modulation appearing in the bulk counting rate from a low-background PPC germanium detector, in a region of the energy spectrum exhibiting an exponential excess of unknown origin. Examined alone, its statistical significance is modest (2.2 $\sigma$ in the basic form of analysis presented here). However, its phase is compatible with that  predicted by halo simulations \cite{purcell,kuhlen2}, and observed by DAMA/LIBRA. When a WIMP origin is considered, the favored dark matter particle mass corresponds to a region of present interest in indirect searches \cite{dan1,aba,dan2,dan3} and numerous particle models \cite{bottino,andreas,model10,model1,model2,model3,model4,model5,model6,model7,model8,model9,model11,model12}.  Most importantly, the large fractional amplitude of the observed modulation would entail a non-Maxwellian component for the local galactic halo.  This property can potentially bring all reported WIMP detector anomalies into agreement. 

We emphasize once more the difficult nature of the search we have described. Presently unknown backgrounds and/or instrumental effects could mimic an annual modulation signature in dark matter detectors. However, a coherent picture may be starting to emerge from present anomalies in direct searches for dark matter: in combination with other results, it should be possible to use the present CoGeNT dataset to constrain the local velocity structure of a hypothetical galactic WIMP halo, an exercise in embryonic ``WIMP astronomy". Such predictions should be testable in the near future, with the advent of precise kinematic information for nearby stars  from the GAIA satellite \cite{kuhlen,gaia}, and more information from other WIMP searches, including the ongoing CoGeNT upgrade, the C-4 experiment \cite{c4}.\\

\section*{Acknowledgements}

We are indebted to SUL personnel for their constant support in operating the CoGeNT detector, and to D. Hooper and N. Weiner for useful input. This work was sponsored by NSF grants PHY-0653605, PHY-1003940, and the PNNL Ultra-Sensitive Nuclear Measurement Initiative LDRD program.  Additional support was received from the Kavli Institute for Cosmological Physics at the University of Chicago through grant NSF PHY-1125897, and an endowment from the Kavli Foundation and its founder Fred Kavli. N.E.F. and T.W.H. acknowledge support by the DOE/NNSA Stewardship Science Graduate Fellowship program (grant number DE-FC52-08NA28752) and the Intelligence Community Postdoctoral Research Fellowship Program.\\

\clearpage


\begin{thebibliography}{9}
\bibitem{note} We have nevertheless remarked in the past a trend for these K-shell EC rates, in combination with L/K EC ratios from \cite{Bahcall}, to overpredict CoGeNT's L-shell EC rates by $\sim$10\% \cite{longcogent,Aal11b}. This is also noticeable in the top panels of Fig.\ 4 and Fig.\ 5. Environmental effects on L-shell electron wave functions are under investigation as a possible cause. Such in-medium effects on the L/K EC ratio have been previously observed \cite{voytas,ray}. Applying a correction to the predicted L-shell EC rate to account for this possible effect produces a negligible change to the best-fit modulation parameters found.
\bibitem{note4}The periodogram algorithm used here \cite{numerical} samples 72 frequencies for the 40 time bins in Fig.\ 5, each with exponentially-distributed power for pure noise \cite{period2}. It is therefore not unexpected to find relatively significant spurious power at random non-annual periods for some of the groups examined in Fig.\ 5, or a finite width to some peaks in the periodogram (only $\sim$40 of these frequencies are independent \cite{numerical}). Separately, we find that the periodogram is not a good indicator of the persistence of a possible modulation over the entire time span examined. For instance, this persistence is noticeable upon simple inspection of the 0.5-2.0 keVee bulk group in Fig.\ 5, whereas power centered around a best-fit $T\!\!=$446 days for the 2.0-4.5 keVee surface group is mostly restricted to the range $\sim$550-1,000 days. Still, a periodogram assigns more significance to the modulation in this latter group (Fig.\ 6), while the likelihood-ratio method estimates it at just 1.5 $\sigma$ (2.2 $\sigma$ for the 0.5-2.0 keVee bulk group).
\bibitem{note2}Earlier \cite{Aal11b} we found a slightly larger significance (2.8$\sigma$) from a live-time exposure a factor 2.5 times smaller than examined here. At first sight, and leaving possible statistical fluctuations aside, this might seem against expectations. However, the much more restrictive rise-time cut ($t_{10-90}<0.7\mu s$) presently applied to extract a subset of "pure" bulk events results in a marginal increase of their statistics in the present cumulative dataset. Namely, the 0.5-2.0 keVee "bulk" sample size in \cite{Aal11b} was composed of 1,766 events, and is now 2,264 events. For the important 0.5-1.0 keVee energy range, where a low-mass WIMP would be expected to dominate, these are 830 and 624 events respectively, i.e., an actual decrease in available statistics. It is therefore not surprising, from a purely statistical point of view, not to have observed an increase in modulation significance. A comparison between the top panels in Fig.\ 5 here and in Fig.\ 4 of \cite{Aal11b} indicates that the similar modulation observed in \cite{Aal11b} was indeed dominated by events with $t_{10-90}<0.7\mu s$.
\bibitem{note3}Decomposing the dataset into smaller pre-fire and post-fire periods, the preference is 1.6 $\sigma$ and 1.2 $\sigma$, respectively. This preference is penalized for the post-fire subset by the fluctuations visible in the top panels of Fig.\ 5 around 1,005 and 1,125 days (no anomalous behavior was observed during those periods). At the time of this writing, we finalize an unbinned maximum-likelihood analysis in three-dimensional space (time, energy, rise-time) able to test different halo models, providing an improved assessment of statistical significance.
\end{thebibliography}
\end{document}